# A Prediction Model of the Project Life-span in Open Source Software Ecosystem

Zhifang Liao[1], Benhong Zhao[1], Shengzong Liu[2,*], Haozhi Jin[1], Dayu He[1], Liu Yang[1], Jinsong Wu[3], Yan Zhang[4]

(1) Department of Software Engineering, Central South University, China
(2) Information Management Department, Hunan University of Finance and Economics, Changsha, Hunan, China
of Engineering and Built Environment, Glasgow Caledonian University
(3) Department of Electrical Engineering, Universidad de Chile, Santiago, Chile
(4) Department of Computer, Communication and Interactive System, School

**Abstract** In nature ecosystems, animal life-spans are determined by genes and some other biological characteristics. Similarly, the software project life-spans are related to some internal or external characteristics. Analyzing the relations between these characteristics and the project life-span, may help developers, investors, and contributors to control the development cycle of the software project. The paper provides an insight on the project life-span for a free open source software ecosystem. The statistical analysis of some project characteristics in GitHub is presented, and we find that the choices of programming languages, the number of files, the label format of the project, and the relevant membership expressions can impact the life-span of a project. Based on these discovered characteristics, we also propose a prediction model to estimate the project life-span in open source software ecosystems. These results may help developers reschedule the project in open source software ecosystem.

## 1 Introduction

Recently, the Software Ecosystems (SECOs) have attracted more and more attentions are in the software engineering research communities. Manikas, Konstantinos, and K. M. Hansen have defined SECOs as "the interaction of a set of actors on top of a common technological platform that results in a number of software solutions or services"[1]. GitHub[2] and SourceForge[3] are typical SECOs. SECOs are becoming are important in continuing development, improving market recognitions, or increasing revenues[4]. Jaap Kabbedijk and Slinger Jansen[5] have exploited social network and statistical analysis techniques to analyze the elements, characteristics, descriptive, roles, cliques and relationships in the open source Ruby ecosystem, and found that only a small percentage of 'core', approximately 10%, of all developers and Ruby packages are dominant in the ecosystem. The relevant development community would benefit from motivating current developers to work together more, rather than supporting new developers to get a healthy ecosystem. So, analyzing SCEOs is of great significance in improving software development.

Researchers try to unify nature ecosystem and software ecosystem, and quote conceptual terms from nature ecosystem to describe SCEOs characteristics, includes ecosystem health, ecosystem diversity and so on. Hansen et al.[4] defined the health of a software ecosystem as the ability of the ecosystem to endure and remain variable and productive over time. They divided SECO Health into three classifications, includes actor health, software health and orchestration health. Slinger Jansen[6] provided the Open Source Ecosystem Health Operationalization, a framework that is used to establish the health of an open source ecosystem. Besides, S. Daniel et al.[7] performed a diversity-related analysis on SourceForge and they found out that diversity in project roles and experiences positively impacts project success. Nicholas Matragkas et al.[8] also analyzed the Biodiversity of GitHub, and showed that the percentage of core developers and active users does not change as the project grows. Some other studies the composition of Software Ecosystems instead of SECOs health and Biodiversity.. P.R.J. Campbell et al.[9] proposed a Three-Dimensional View of SECOs, it contains Business, Architecture and Social dimensions, which are closely integrated through software engineering processes. Deguang Zhang et. al..[10] proposed Open Source Software Ecosystems based Technology platform, Actor and software projects. Besides, some researchers applied the above-mentioned theories to analyze concrete Software Ecosystem. Rick Hoving et al.[11] analyzed the characteristics of Python Software Ecosystem in GitHub, and found that the FOSSE of Python is growing rapidly, and that of each year is better than that of the pevious year. Similarly, we also can utilize these theories to analyze other SECOs to improve the software development.

Inspired by nature ecosystems, this paper studies the project life-span in Software Ecosystems. We analyze the internal or external characteristics of free open source projects to find out which characteristics can be used to estimate the life-span length of existing projects. This paper is organized as follows. Section 2 provides a list of research questions and introduces the concept of data gathering. Section 3 defines the life-span of open source project and gives an overview on the life-span of whole GitHub SECO. Section 4 explores how the project characteristics reflect or influence the life-span of projects. Section 5 presents a prediction model for the life-span of projects in SECO and provides a analysis on it. The last section discusses the findings and concludes the paper.

## 2 Research questions and data gathering

The goal of the paper is to study the project life-span in open source software ecosystems. The research questions the paper answers are listed as follows:

**Q1** *What is the definition of project life-span in open source SECOs?*

**Q2** *How long are the life-span of projects in whole open source SECOs?*

Before our research, we clarify the relevant definition for our study. These two above-questions would be answered in section 3.

**Q3** *Which characteristics would be related to the project life-span ?*

**Q4** *What are the relations between characteristics and the project life-span ?*

We would like to explore whether there exist some relations between characteristics of SECOs and the project life-span, which would be answered in Section 4.

**Q5** *how to estimate the project life-span in open source software ecosystems ?*

We will present a prediction model based on the investigation result of Q5, and the performance of model would be shown in Section 5.

Our dataset downloads from GHTorrent, a website for gathering GitHub data. The dataset stores the project data before October 30,2013, which include data from 5342633 projects and 2437234 users. We collect the experimental data follow the rules:

(1) the projects have no commitments after April 30,2013, to ensure that they have no commitments in six months.

(2) the projects are original works and have not forked from other repositories.

(3) the projects have not been deleted.

Based on these rules, we obtain an experimental dataset that contain 843763 projects.

## 3 Defination of software project life-span

In biological world, every life has his own life-span, for example, healthy elephant tortoise can live for 300 years, drosophila can only live for three days, and as normal humans, our life-span usually is 60-90 years. Life grows constantly from birth to death. Inspired by the nature ecosystem, we find the similar life stage of the projects in open source software ecosystem. Table 1 shows the life stage of project and a comparison with humans.

To ensure that the reproduction of your illustrations is of a reasonable quality, we advise against the use of shading. The contrast should be as pronounced as possible.The human life-span is the living period from birth to death, therefore, the project life-span is defined as the period from birth to death. In other words, the project life-span is the period from the time that the project is created to the time have no commitment any more. For example, a project was created at May 5, 2015, and it died at May 29, 2015, then its life-span is 24 days. We analyze the dataset GitHub2013, and obtain the statistics of all projects life-span whose death time was before '2013/4/30 23:58:59', the result as Fig 1 show.

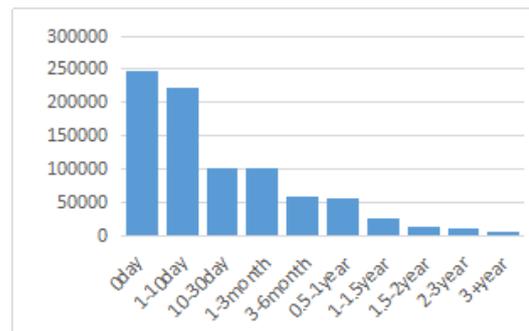

**Fig. 1.** The numbers of projects in different life-span segments

The vast majority of projects only lived for less than 10 days. Almost 250 thousands projects have not lived for less than or equal one day, and about 225 thousands lived for 1 to 10 days. Most of these projects were committed several times after repositories creation , and then were abaendoned. They were usually test projects for learning how to use the GitHub system, or they were some failed projects. The numbers of projects that had lived for 10-30 days and 1-3 months are both up to 100 thousands. These projects are always small, and have lower complexity and less works to be completed. Most of them are developed by users who are amateurish and always used for personal purposes. For the longer life-span, the number of projects decreases, in other words, less projects live for longer. Only 4627 projects lived for more than 3 years, most of them were completed by professional or skillful developers, always released several versions, and had continual improvement in functions or user interfaces.

**Table 1.** The life stages of software project and humans

| Life stage | Software project | Humans |
|---|---|---|
| born | The time when the software project is created in SECO repository | The moment when the human being comes to the world |
| adult | The beta version is released and start providing service to users | One person reaches 16 or 18 years old, and starts independent living, taking responsibility for himself or herself |
| middle age | the time from second version to later versions when the project attains stability and fault tolerance and so on. | One person becomes mature, has enough ability to handle the things in life or work |
| die | has no code or comment committed any more | stop breath and heartbeat |

## 4   Analysis

In nature ecosystems, the species life-span is determined by genes and their biological features. For example, larger animals live longer, while for those reaching sexual maturity earlier, they will live shorter. Inspired by the biological laws, we try to study similar laws about project life-span in open source SECOs. Firstly, according to the enterprise projects or non-open source software, the life cycle is closely related the size of projects and developers employed. As a reference, we select project description, the file number and membership of projects as the characteristics that are likely to influence the project life-span in open source SECOs. In addition, the programming language and project label are taken into consideration as well. The discussions about the relation of projects life-span with this four characteristics are provided as follows. Our statistical analysis is based on the projects mentioned in section 3. The life-span of these projects lasted longer than 10 days, to ensure that they were natural death. In other words, these projects have been normally completed.

### 4.1   Project description

Project description always contains the statements of the project functions, features or operation instruction. So Project description implies the software size and the degree of complicacy and difficulty in software development. In GitHub, developers write down the description in the Readme file. Figure 2 shows the screen-shot of Readme webpage.

We crawled the projects homepage in GitHub, and count the amount of words in Readme contents. And then, explore the relation between the amount of description and project life-span. Figure 3 shows the average length of life-span in different amounts of description words.

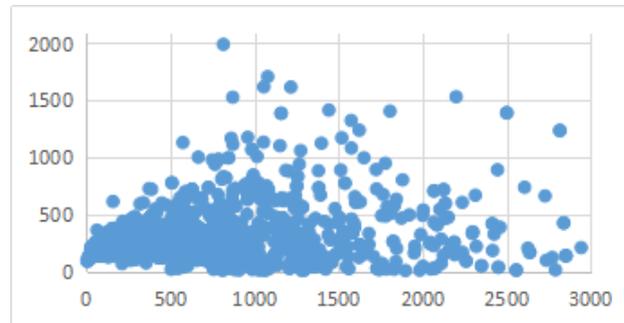

**Fig. 3.** The average length of life-span in different amounts of description words

In figure 3, horizontal axis is the amounts of description words, namely text length, vertical axis is life-span. From the figure, the lengths of most projects descriptions are under 1000 words, and their life-span are under 800 days. When it exceeds 1000 words, the project life-span become discrete and uncertain. We calculate the Pearson correlation coefficient $r$ to evaluate the degree of correlation between this two variables. The $r$ of description length from 0 to 500 is 0.289, that means life-span and description length have positive correlation on them, but the correlation is very weak . The $r$ (500-1000 words) is 0.084, and the $r$ (more than 1000 words) is almost zero, so this two variables have no correlation when the description length exceed 500 words. According to the analysis, we obtian the first conclusion as follow:

*Result 1:* the life-span of project is not relevant to the length of project description.

### 4.2   Programming language

When a software project starts, the programming language is one of the vital factors to be considered. Different languages are suitable for different application scenarios. For example, Python is usually applied to scientific calculation, Java is applied to website development. The choices of programming languages for developers depend on the features of projects. Figure 4 shows the usage of the top 12 programming languages in GitHub, where JavaScript is the most popular programming language, about 25% of projects use it. The popularity of Ruby, 14.64%, is close to that of Java, reaching14.24%. However, only 2.31% of projects use Perl as the chosen programming language. There are huge differences in the usage of programming languages.

Through the analysis of projects in different programming languages, we find that it exists huge differences in the life-span of them as well. Table 2 shows the statistics of the life-span in different programming languages. From the table, Java projects ha-

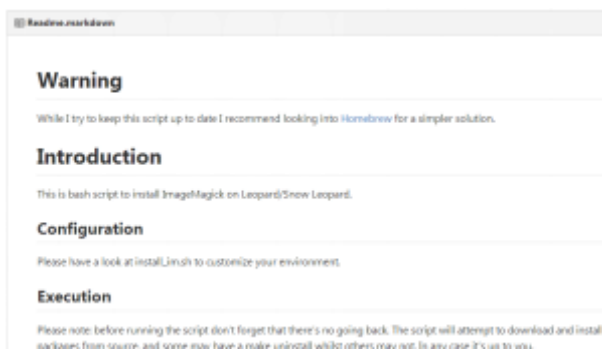

**Fig. 2.** A Readme webpage of project

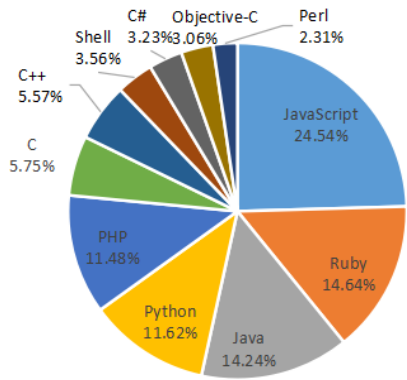

**Fig. 4.** The usage of programming languages

ve shortest average life-span, only 145.6 days, about 5 months. C# and JavaScript projects have short average life-span as well. However, Perl projects achieve the longest average life-span, reach 343 days, almost one years, and more than twice as long as the average life-span of Java projects. The projects based on different programming languages lasted quite different average life-spans, and and we can find the same relative difference to the other statistics parameters, such as first quartile(Q1), median, and third quartile(Q3). The Q1 of Java project life-span is only 25 days, median is 63 days, and Q3 is 182 days, that means 1/4 of java projects, their life-span are under 25days, 1/4 are upon 182 days, and half are under 63 days. These parameters of Java projects are all low. On the contrary, the parameters of Perl projects are all high, especially the Q3 reach 526 days. Therefore, our discoveries about the relations in programming language and life-span are listed as follows:

*Result 2*: the life-span of software projects in Open Source Software Ecosystems is closely related to the chosen programming language.

**Table 2.** The statistics of life-span in different programming languages.

| Language | average | Q1 | median | Q3 |
|---|---|---|---|---|
| Java | 145.6598 | 25 | 63 | 182 |
| C# | 154.2792 | 26 | 72 | 195 |
| JavaScript | 160.0089 | 23 | 79 | 128 |
| Objective-C | 167.9306 | 24 | 67 | 213 |
| C++ | 169.6528 | 28 | 76 | 215 |
| PHP | 182.7401 | 32 | 93 | 250 |
| C | 206.0435 | 31 | 90 | 273 |
| Python | 210.2633 | 34 | 105 | 289 |
| Ruby | 213.5365 | 27 | 81 | 272 |
| Shell | 237.9406 | 45 | 137 | 300 |
| Perl | 343.0235 | 58 | 211 | 526 |

As we know, each programming language has its own advantages or disadvantages, and may be suitable for different scenarios. A lot of development tools and frameworks are provided by software venders and providers for some mature and popular programming languages, such as java, C#. . Thus, projects using these programming languages can be completed in short time. On the other hand, some less popular programming languages, such as Perl and Shell, are relatively less known by developers, with less contributions, and the relevant projects experienced longer life-spans. The above explain our result 2.

### 4.3 File number

In natural ecosystems, the genes of species have already coded and constrained the birth and deadth time for animals, before the animals come into the world. Thus, we know that the sea turtle usually could live for hundred years. In software ecosystem, we think that the size of projects determine the length of projects life-span. In this paper, we use the number of files to measure the size of project. If a project has more files stored in the repository, including source files, configuration files and others, the project would have a more larger size. In order to explore the relation of project life-span and the project size, we have used web crawlers to extract the file number of projects from the Github website. Figure 5 shows the life-span of projects with different file numbers.

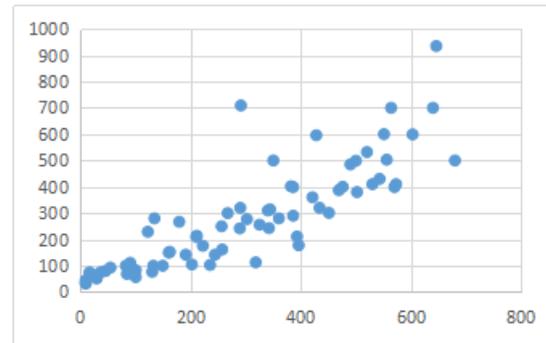

**Fig. 5.** The life-span with different file number.horizontal axis is the file number in the projects, vertical axis is average life-span of projects in different file number.

Obviously, when the projects require more files in the repositories, they are likely to have longer life-spans. From the figure 5, the projects with less than 200 files usually live for less than 200 days. On the contrary, if projects has more than 400 files, their life-span will up to more than 300days. We calculate the Pearson correlation coefficient *r* to evaluate the degree of correlation between this two variables, and the value of *r* is 0.85, that means the life-span has a strong correlation with the file number of a project. Therefore, we obtain the second conclusion result as follow:

*Result 3*: the project life-span would increase with the increase of file number.

### 4.4 Project membership

The completion of software is not only related to the internal characteristics of projects, but also strongly relevant to the number, quality, and capacities of the software developers. Larger projects always usually require more developers. In GitHub, the repository owner set other active users as collaborators, who consist of core developers of the project together. Table 3 shows the numbers of projects that have different quantities of core developers. The vast majority of projects have only one core developer, who is the repository owner himself. These projects are always individual and small, and do not require more developers. Less projects has 2 or more core developers. Besides, other non-core developers could contribute to the projects by committing PullRequest, a special method for them to submit codes. Here we only consider the core developers.

To explore the relation of life-span and projects membership, we analyze the quantity and quality of core developers both. Here we use number of followers to core developers to measure the quality of core developers. If a developer has higher quality, he or she would attract more other users to follow him or her. According to different chosen programming languages, the life-spans with project memberships were classified into 3 groups. The results as follow scatter diagram shows. In Figure 6, the horizontal axis expresses the numbers of core developers or followers, while the vertical axis is life-span. The left figures describes the relation between the number of core developers and the life-span of projects in Java, JavaSc-

**Table 3.** The number of projects that have different numbers of core developers.

| quantity of core developers | The number of projects |
| --- | --- |
| 1 | 260840 |
| 2 | 17939 |
| 3 | 6359 |
| 4 | 3157 |
| 5 | 1801 |
| 6 | 966 |
| 7 | 668 |
| 8 | 466 |
| 9 | 481 |
| 10 | 294 |

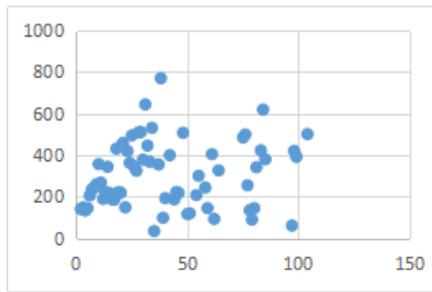 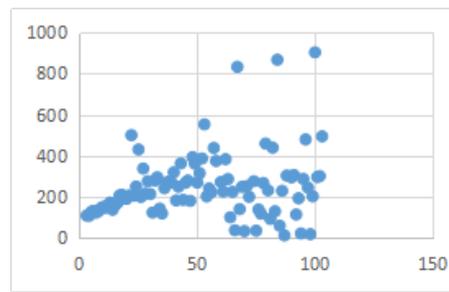

**Fig. 6.1** Left is the core developer amounts and life-span, right is figure of the follower amounts and life-span, They both are in Java project

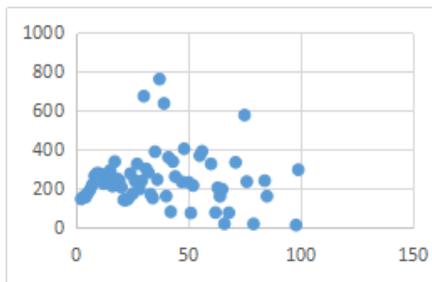 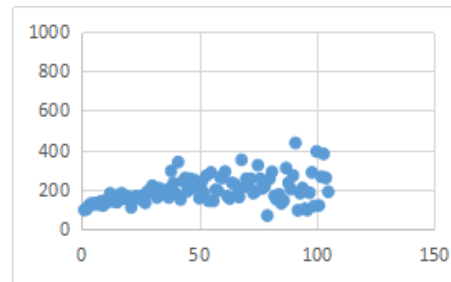

**Fig. 6.2** Left is the core developer amounts and life-span, right is figure of the follower amounts and life-span, They both are in JavaScript project

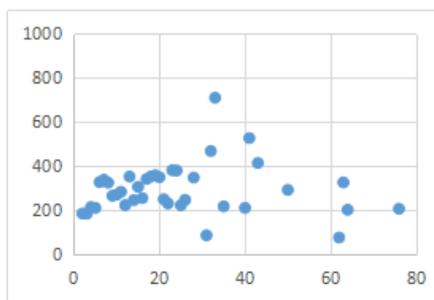 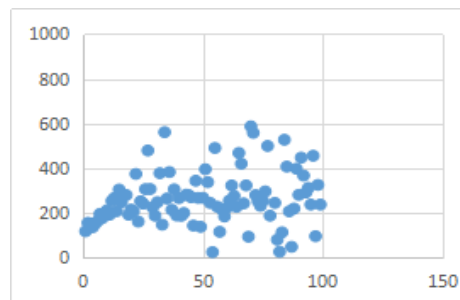

**Fig. 6.3** Left is the core developer amounts and life-span, right is figure of the follower amounts and life-span, They both are in PHP project

ript and PHP languages respectively. When numbers of core developers are under 20 users, we can find smaller increase trend of life-spans, but when the numbers are more than 20, the life-spans show diversity and can not find any regularity. The right figures describe the number of followers and life-span. As the figures show, the data is aggregate obviously in these three languages both, compared with left figures. Importantly, there is an obvious positive correlation between the number of followers and life-span, especially for the choice of JavaScript language. We have calculated the Pearson correlation coefficient r of followers amounts and life-span as well, and the r of Java is 0.3852, the r of JavaScript is 0.6338, the r of PHP is 0.2924, where the coefficient r reflects the positive correlation, and much closer to 1, much obvious the correlation is stronger. According to our analysis, we can draw conclusions as follows:

*Result 4:* the life-span of software projects in Open Source Software Ecosystems is independent of the quantity of core developers.

*Result 5:* there is a positive correlation between the life-span of projects and the quality of core developers in Open Source Software Ecosystems. If the core developers have more followers in total, the life-span of projects would be longer.

In open source software ecosystems, one capable developer more possibly become the core of a complex project, and more complex the project is, more time would be needed for the project completion. That explains result 5.

### 4.5 Project label

Each project has its own features in Open Source Ecosystem, it may belongs to "Database" or has a great "compatibility". These features would be marked on the projects, and as a label of projects. This helps the project spreading in the GitHub. Different features imply different difficulties for project completion. For example, the developers expect a good effort to make the project obtain "compatibility", that may result in longer life-span.

**Table 4.** The statistic of project label and life-span.

| Label | Life-span | Label | Life-span |
|---|---|---|---|
| editor | 577.0 | server | 299.5 |
| Linux | 551.1 | model | 297.0 |
| Compatibility | 521.7 | IOS | 260.5 |
| optimization | 503.5 | build | 259.1 |
| template | 493.4 | architecture | 252.3 |
| Windows | 474.9 | web | 241.4 |
| Website | 463.9 | bug | 212.5 |
| security | 413.9 | Maps | 172.5 |
| enhancements | 395.0 | data IO | 126.0 |
| Mobile | 389.0 | back end | 124.5 |
| API | 370.6 | J2ME | 70.0 |
| Database | 355.8 | HTML 5 | 70.0 |
| plugin | 318.5 | bootstrap | 60.0 |

To explore the relation between project label and life-span, we make a statistical analysis, and the relevant results are shown in Table 4.

From Table 4, the projects with "editor" label have the longest average life-span, up to 577.0 days. The projects with "Linux" or "Compatibility" label also have long life-spans. However "J2ME", and "HTML" or "bootstrap" projects have only 60.0 to 70.0 days of life-spans, far less than other projects, and they usually are web projects and can find lots of mature and stable development frameworks for software development. Thus, the development of those projects are simple and short-term. Differently, projects with "Compatibility" or "Linux" are difficult for ordinary developers, and they require more professional knowledge to achieve the project goals, which lead to longer life-spans. Therefore, we can conclude another result as follows:

*Result 6:* the life-span of software projects in Open Source Software Ecosystems is closely related to the project label.

## 5 Prediction model

From the above analysis, the projects life-span in open source software ecosystems is determined by the file number, and related to the chosen programming language, the membership and the project label. Based on these findings, we present a prediction model to estimate the project life-span. In our work, *LP(p)* is used to represented and predict the life-span of the project *p*, and can be calculated as follows:

$$LP(p) = \alpha \bullet \log_2[n(p)] \bullet l(p) \bullet \log_2[m(p)] + \beta \bullet lab(p) \quad (1)$$

where *n(p)* is the file number of *p*, *l(p)* is the influence of the chosen programming language and the *m(p)* is the impact of membership, which, in this paper we use the followers number of core developers in the project p to represent it. The *lab(p)* is the influence of project label. *α* is a factor, here it represents the time length to complete one project file, *β* is used to adjust the effect of project label. This model is consistent with our findings.

Different from traditional software development, the open source software projects are usually developed by the developers during their spare time, so the works for projects are typically discontinuous, which would seriously affect the prediction result of the model we have presented above. Figure 5 shows two projects that have different types of contributions. The green areas mean it have commits to the project. From the figure 5(a), we can see the work of the project was continuous during the period of from the beginning of 2012 to the end of 2013. On the contrary, the project in figure 5(b) has stoped working for several times and it only worked for about two month. In the terms of our prediction model, it is difficult to estimate the life-span of project in this type.

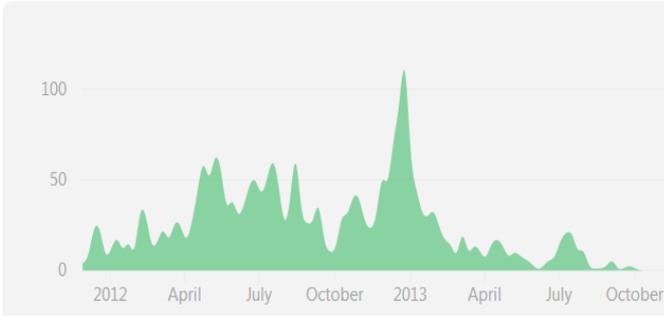

**Fig. 5(a)** Project for continuous contributions

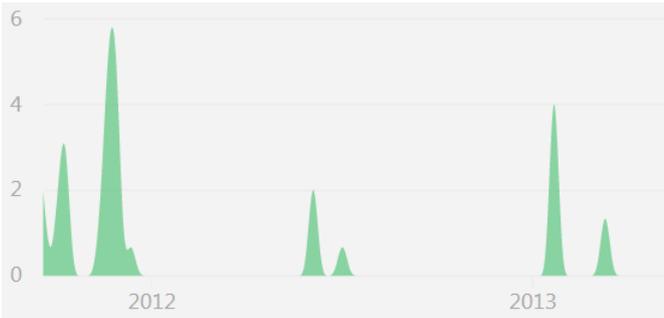

**Fig. 5(b)** Project for discontinuous contributions

Taking this above-mentioned problem into consideration, we introduce the parameter, non-working ratio, in this paper, which is defined as the ratio of the non-working days in the whole life-span. $r(p)$ is the non-working ratio of project p, which is expressed as follows: where $d$ is the non-working days, and $w$ is the length of

$$r(p) = \frac{d}{w} \qquad (2)$$

real life-span of p. In our work, we decide whether a day is a non-working day according to its commit records to the project. If the project has no commits in the day, the day is considered as a non-working day. Due to this special feature of open source software ecosystem, we only consider the cases that the number of continuous non-working days is more than 6. The code that computes the value of $r(p)$ is written as follows:

```
Def get_nowork_ratio(commit_dates,life_pan):
    data_lenth = len(commit_dates)
    no_work_day = 0
    if data_lenth >= 2:
        for i in range(1, data_lenth):
            dl = (commit_dates[i] -
                  commit_dates[i-1]).days
            if dl > 6:
                free_day = no_work_day + dl
    return no_work_day/float(life_pan)
```

Then, we verify the validity of the prediction model. In our work, we take $α$ equal 1.204, and the value is the average time to complete a project file. The variable $l(p)$ takes the java programming language as baseline. If $p$ is a java project, the value of $l(p)$ is 1, and if $p$ is a C# project, the value of $l(p)$ is 1.059, according the ratio of average life-span in different chosen progamming languages. $β$ is set to 0.8 as a adjustment factor. $lab(p)$ is the value of the average life-span with different labels subtracting the average life-span of all projects. For example, the average life-span of projects with "Maps" is 172.5 days, and the average life-span of all projects in dataset is 149.4, then the $lab(p)$ is 23.1. We select the relative error between the predicted and the real life-span as the metrics of model, and choose the projects that non-working ratio $r(p)$ less than 0.3 as the experimental data. The prediction results are shown in Figure 6.

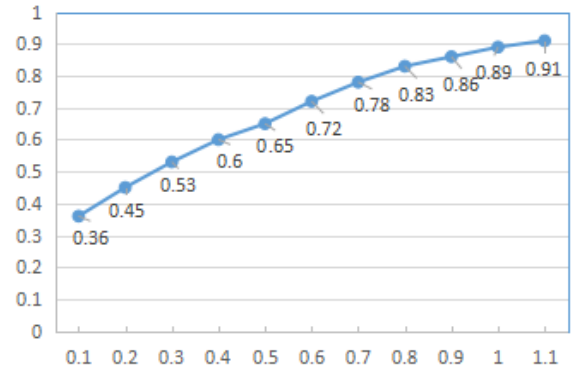

**Fig. 6** The ratio of projects in different relative error

In Figure 6, the horizontal axis is the value of relative error, the vertical axis is the ratio of projects whose relative error less than $x$. From the Figure 6, 36% of projects, their relative errors of prediction result are less than 0.1, and 53% of projects, the relative errors of them are less than 0.3. So, our prediction model is effective and valid..

*Result 6:* we can use the prediction model we presented above to estimate the life-span of software projects in Open Source Software Ecosystems with the low non-working ratio limitation.

## 6   Discussion and conclusion

The main work of this paper is to explore the life-span of projects in free open source Software Ecosystem. Our work is inspired by the biological laws about the animals life-span in nature ecosystems. Firstly, we have used the life stage of human as a reference, and proposed a definition of project life-span. The answer to the research question 1 is (1) project life-span is the time from the project being created to having no commit any more. Secondly, we have made a statistical analysis on the life-span of projects in GitHub, and provided the answer to research question 2: (2) the vast majority of projects only lived for less than 10 days, and there still exist lots of projects lived for 10 days to 3 months, but less projects lived for longer. Thirdly, we have analysed the characteristics of projects in GitHub, and find that (3) the project life-span is related to the file number of projects, programming language, project membership and label. This explains research question 3.

More over, we have made a detailed analysis on these internal or external characteristics, and provided the answer to question 4. The answer includes 5 results which we have concluded in section 4. (4)(a) there is a strong positive correlation between project life-span and the file number of project. (b) project life-span is closely related to the chosen programming language. As we have observed, the Java and C# projects have short life-spans, only about five months, but Perl projects lived for about one year on average. (c) project life-span is independent of the quantity of core developers, but (d) related to the quality of core developers. In the paper, we have used the number of followers to represent the quality of core developers, and find that the project with more followers of the core developers always has a longer life-span. (e) the project label imply the length of project life-span. we have gathered the labels of the projects, and the results have shown large differences on the life-span of the projects with different labels. For example, the projects with "editor" label have an average life-span of 577 days, but the projects with "HTML 5" have only 70 days.

Based on the above findings, we have presented a prediction model to estimate the project life-span in open source software ecosystems. The model relies on those characteristics and is consistent with our findings. With the low on-working ratio limitation, our model has shown its validity. This explains research question 5.

This paper has presented the findings based on the dataset of GitHub, a typical free open source software ecosystem. Those findings can help developers make the proper time schedule of projects in open source SECOs. For example, if someone wants to finish a software project in a short time, he or she should choose a appropriate programming language, like Java or C#, and if he or she wants to start a project about "Database" in open source SECOs, he needs to prepare for long time effort, about one year. In this paper, we only have explored 5 features of the open source projects that related to the life-span. In the future, more factors about the life-span would be studied in the SECOs field. In addition, it is worthwhile studying how to improve the performance of prediction model on the projects life-span that we have presented in this paper.